\documentclass[11pt]{article}
\usepackage{amsmath,amssymb,amsthm,mathrsfs}
\usepackage{geometry}
\usepackage{tikz}
\usetikzlibrary{decorations.pathmorphing,positioning}
\usepackage[utf8]{inputenc}
\usepackage{hyperref}
\usepackage{enumitem}
\usepackage{cite}
\usepackage{setspace}
\AtBeginDocument{\setstretch{1.2}}
\usepackage[T1]{fontenc}
\usepackage{amsmath,amssymb,mathtools}
\usepackage{kpfonts}

\hypersetup{
	unicode=true,            
	pdftoolbar=false,       
	pdfmenubar=false,        
	pdffitwindow=false,      
	pdfstartview={FitH},     
	pdftitle={},             
	pdfauthor={},            
	pdfsubject={},           
	pdfcreator={},           
	pdfproducer={},          
	pdfkeywords={},          
	pdfnewwindow=true,       
	colorlinks=true,         
	linkcolor=blue,          
	citecolor=magenta,       
	filecolor=green,         
	urlcolor=blue            
}

\newtheorem{Prop}{Proposition}[section]

\allowdisplaybreaks
\numberwithin{equation}{section}

\title{Small-$b$ expansion of the DOZZ formula for light operators}

\author{
Franco Ferrari$^{a}$\footnote{e-mail: franco.ferrari@usz.edu.pl}
$\quad$
Marcin R.~Pi\c{a}tek$^{a}$\footnote{e-mail: marcin.piatek@usz.edu.pl}
$\quad$
Artur R.~Pietrykowski$^{a}$\footnote{e-mail: Hearthie@gmail.com}
\\[8pt]
{\normalsize ${}^{a}$Institute of Physics, University of Szczecin,}\\ 
{\normalsize         Wielkopolska 15, 70--451 Szczecin, Poland}
}

\date{}

\begin{document}
\maketitle

\begin{abstract}
We present a systematic asymptotic small-$b$ expansion of the
Liouville DOZZ three-point structure constant in the light-operator
regime $\alpha_i=b\sigma_i$ in the limit $b\to0$.
In this limit the DOZZ structure constant admits the asymptotic
representation
\[
C(b\sigma_1,b\sigma_2,b\sigma_3)
\sim
{\cal P}(b;\sigma_i)
\left[
1+\sum_{n\ge1}b^{2n}\Omega_n(\sigma_1,\sigma_2,\sigma_3)
\right],
\qquad b\to0 ,
\]
where ${\cal P}(b;\sigma_i)$ is an explicitly determined prefactor.
Using Thorn's asymptotic expansion of the $\Upsilon_b$-function,
we derive explicit closed-form expressions for the first coefficients
$\Omega_n(\sigma_i)$ and show that they form a symmetric polynomial
structure in the variables $\sigma_i$.
The resulting asymptotic expansion provides explicit finite-$b$ corrections
to the leading light-limit Liouville three-point data and
furnishes a practical framework for applications in celestial
holography. In particular, it provides Liouville data relevant for a
perturbative program aimed at studying finite-$b$ corrections to
celestial Yang--Mills amplitudes.
Finally, we formulate such a perturbative Liouville framework and
discuss several directions for its future development.
\end{abstract}

\newpage
\tableofcontents

\section{Introduction}
Let \(\Sigma_g\) be a compact Riemann surface of genus \(g\) endowed with a reference metric \(g_{ab}\) 
and scalar curvature \(R\). Liouville theory describes a field \(\varphi\) governed by the action:
\begin{equation}\label{SLg}
S_L[\varphi;g]
=\frac{1}{4\pi}\int_{\Sigma_g}\!\mathrm{d}^2x\,\sqrt{g}\,
\Bigl(g^{ab}\partial_a\varphi\partial_b\varphi
+Q R\,\varphi+4\pi\mu\,\mathrm{e}^{2b\varphi}\Bigr),
\qquad Q=b+\frac{1}{b},
\end{equation}
where \(b\) is the Liouville coupling and \(\mu\) the cosmological constant. 
In the genus-zero case, Liouville theory on the Riemann sphere is locally described by the Lagrangian density
\begin{equation}
\mathcal{L}=\frac{1}{2\pi}\!\left(\partial\varphi\bar\partial\varphi+\pi\mu\,{\rm e}^{2b\varphi}\right),
\end{equation}
whose Euler--Lagrange equation is the Liouville equation
\begin{equation}
\partial_z\partial_{\bar z}\,\varphi(z,\bar z)
=
\pi\mu b\,{\rm e}^{2b\varphi(z,\bar z)}.
\end{equation}
Here \(z\) denotes the local complex coordinate on the plane and 
\(\partial\equiv\partial_z=\partial/\partial z,\ \bar\partial\equiv\partial_{\bar z}=\partial/\partial\bar z\). 

Upon quantization, Liouville theory becomes a nonrational two-dimensional conformal field theory 
with central charge $c=1+6Q^2$. The spectrum of primary fields is continuous and is spanned by the exponential 
vertex operators
$\mathsf{V}_\alpha(z,\bar z)={\rm e}^{2\alpha\varphi(z,\bar z)}$,
with labels
$\alpha\in\mathbb{S}=\tfrac{Q}{2}+i\mathbb{R}_{\ge0}$,
and identical holomorphic and antiholomorphic conformal weights
$h(\alpha)=\bar h(\alpha)=\alpha(Q-\alpha)$.\footnote{For further details, see Appendix~\ref{app1} and 
refs.~\cite{BPZ,S,DO,T1,ZZ5,T3,T4,Nakayama:2004vk}.}

Conformal symmetry fixes Liouville three-point correlators up to a single structure constant,
given by the Dorn--Otto--Zamolodchikov--Zamolodchikov (DOZZ) formula:
\begin{eqnarray}
\label{DOZZ}
C(\alpha_1, \alpha_2, \alpha_3)
&=&
\left[\pi\mu\gamma(b^2)b^{2-2b^2}\right]^{(Q-\alpha_1-\alpha_2-\alpha_3)/b} \times\nonumber
\\[5pt]
&&\hspace{-60pt}\frac{\Upsilon_{0}\Upsilon_{b}(2\alpha_1)\Upsilon_{b}(2\alpha_2)\Upsilon_{b}(2\alpha_3)}
{\Upsilon_{b}(\sum_{i} \alpha_i - Q)\Upsilon_{b}(\alpha_1 + \alpha_2 -
\alpha_3) \Upsilon_{b}(\alpha_2 + \alpha_3 - \alpha_1)\Upsilon_{b}(\alpha_3
+\alpha_1-\alpha_2)}.
\end{eqnarray}
The structure constant~\eqref{DOZZ} is expressed in terms of the special function
$\Upsilon_b(x)$, which can be constructed from Barnes' double gamma function.
Here $\gamma(x)\!=\!\Gamma(x)/\Gamma(1-x)$, and 
$\Upsilon_0\equiv\Upsilon_b'(0)=\left.\frac{\mathrm{d}}{\mathrm{d}x}\Upsilon_b(x)\right|_{x=0}$.
The DOZZ formula was originally conjectured independently by Dorn and Otto~\cite{DO}
and by Zamolodchikov and Zamolodchikov~\cite{ZZ5}. Systematic derivations were later
provided by Teschner \cite{T4,T1}.

Many modern applications --- most notably those relating Liouville CFT data to
higher dimensional scattering observables in celestial holography \cite{Stieberger2023a,Stieberger2023b,Melton:2024akx}
--- require controlled expansions of exact Liouville ingredients
in regimes where the central charge is large. The semiclassical limit \(b\to0\)
(equivalently \(c\to\infty\)) admits two complementary regimes. In the
\emph{heavy} limit, \(\alpha_i\sim b^{-1}\), correlation functions exponentiate and are governed by the
classical Liouville action on the punctured sphere (see, e.g.,~\cite{HJ04,HJP05}).
By contrast, in the \emph{light} limit $\alpha_i=b\sigma_i$ with fixed
$\sigma_i$, the DOZZ structure constant does not exponentiate in the heavy
sense; instead, it admits a systematic asymptotic expansion in powers of
$b^2$ obtained from the small-$b$ asymptotics of the special function
$\Upsilon_b(x)$.\footnote{The light operators considered below require the standard
	analytic continuation interpretation. For fixed \(\sigma_i\), the light
	momenta \(\alpha_i=b\sigma_i\) do not belong to the physical normalizable
	spectrum \(\mathbb{S}=\frac{Q}{2}+i\mathbb{R}_{\geq0}\). Accordingly,
	throughout this paper, the Liouville structure constant with light
	insertions is understood via the meromorphic continuation of the DOZZ
	formula in the momenta \(\alpha_i\) from the physical spectrum to the values
	\(\alpha_i=b\sigma_i\) required by the light limit, away from its poles.
	The corresponding light exponential operators should therefore not be
	interpreted as normalizable states in the physical Liouville Hilbert space.}

In this paper, we develop a systematic perturbative Liouville approach centered on the light limit
and present a closed-form small-\(b\) expansion of the DOZZ three-point structure constant.
More precisely, we show that for fixed generic values of $\sigma_i$, the DOZZ structure
constant possesses the following small-$b$ asymptotic representation:
\begin{equation}\label{dozzlight}
C\bigl(b\sigma_1,b\sigma_2,b\sigma_3\bigr)
\sim
\mathcal{P}\bigl(b;\sigma_1,\sigma_2,\sigma_3\bigr)
\left[1+\sum_{n\geqslant1}b^{2n}\Omega_n(\sigma_1,\sigma_2,\sigma_3)\right],
\qquad b\to 0,
\end{equation}
where the explicit prefactor \(\mathcal{P}(b;\sigma_i)\) is given by\footnote{Here
\(\pi\tilde\mu\gamma(b^{-2})=\big(\pi\mu\gamma(b^{2})\big)^{b^{-2}}\).}
\begin{align}\label{Pre}
\mathcal{P}(b;\sigma_1,\sigma_2,\sigma_3) &= 
\frac{\pi\tilde\mu\,\gamma(b^{-2})
\left(\pi\mu\,\gamma(b^2)b^{-2b^2}\right)^{1-\sum_{i}\sigma_i}
\gamma\Big(\sum_{i}\sigma_i-1-\frac{1}{b^2}\Big)}{b^{5}}\nonumber\\[5pt]
&\quad\times
\frac{\Gamma(\sigma_{1}+\sigma_{2}+\sigma_{3}-1)\,
\Gamma(\sigma_{2}+\sigma_{3}-\sigma_{1})\,
\Gamma(\sigma_{3}+\sigma_{1}-\sigma_{2})\,
\Gamma(\sigma_{1}+\sigma_{2}-\sigma_{3})}
{\Gamma(2\sigma_{1})\Gamma(2\sigma_{2})\Gamma(2\sigma_{3})}\nonumber\\[5pt]
&\quad\times
b^{-2b^2(\sigma_1+\sigma_2+\sigma_3-1)}\;,
\end{align}
and where the functions $\Omega_n(\sigma_i)$ denote the coefficients of
the resulting asymptotic expansion and encode the finite-order corrections
in the small-$b$ regime. 
Using Thorn's small-\(b\) expansion of the special 
function \(\Upsilon_b(x)\) as a technical tool (see \cite{Thorn}), we obtain closed-form analytic expressions for the first few 
coefficients \(\Omega_n(\sigma_i)\). We verify that these coefficients obey the expected permutation symmetry and we discuss 
their analytic structure as functions of the \(\sigma_i\).

Two conceptual points motivate and frame our calculation. 
First, the leading term in~\eqref{dozzlight} admits a clear minisuperspace
(zero-mode) interpretation in terms of matrix elements of exponential
Liouville vertex operators between Liouville quantum-mechanical states.
The coefficients $\Omega_n$ provide successive finite-$b$ corrections to
the leading light-limit contribution.
Second, and of direct practical importance, the finite-\(b\) coefficients furnish explicit and controlled input 
data for a perturbative bootstrap around the solvable minisuperspace point. 

Expanding four-point functions order-by-order in \(b^2\) and imposing crossing symmetry yields linear relations 
that link corrections to the three-point data, to the spectral measure, and to \(1/c\) corrections of the \emph{global} 
conformal blocks \cite{Perlmutter:2015,Bombini:2018jrg}. 
Because we provide closed-form expressions for the \(\Omega_n\), these otherwise formal 
relations become explicit and testable equations --- for example, furnishing concrete consistency conditions for 
any perturbative (loop) reconstruction based on Liouville input, including proposals relating Liouville data to 
celestial amplitudes \cite{Stieberger2023a,Stieberger2023b,Melton:2024akx}.

The structure of the paper is as follows. In Section~\ref{smallb} we review the derivation of the asymptotic expansion of the 
\(\Upsilon_b\)-function and develop the small-\(b\) expansion of the DOZZ three-point function. 
Section~\ref{perturbative-liouville} formulates the perturbative Liouville program, including functional and spectral methods for 
computing \(\Omega_n\) and the perturbative bootstrap constraints, and discusses its applications to celestial holography. 
We conclude in Section~\ref{conclusions}.

\section{Small-${\boldsymbol b}$ asymptotic expansion of the DOZZ formula}
\label{smallb}
 In this section, we develop a systematic weak-coupling ($b\to0$)
asymptotic expansion of the DOZZ three-point structure constant.
Our strategy proceeds in two steps. First, we derive the small-$b$ asymptotics of 
the special function $\Upsilon_b(b\sigma)$. This step follows the analysis of Thorn~\cite{Thorn}, 
with all intermediate steps provided here for completeness. Second, we apply this expansion to the 
full DOZZ expression to obtain an explicit expansion of the structure constant with light insertions, 
i.e., valid for $\alpha_i = b\sigma_i$ with fixed $\sigma_i$ as $b\to 0$. 
This procedure yields closed-form expressions for the coefficients $\Omega_n(\sigma_i)$ of 
the resulting asymptotic expansion.

\subsection{Asymptotic expansion of the $\Upsilon_{b}$-function}
The $\Upsilon_b$-function satisfies the following functional equations:
\begin{eqnarray}\label{U1}
\Upsilon_b(x+b)&=&\frac{\Gamma(bx)}{\Gamma(1-bx)}b^{1-2bx}\Upsilon_b(x),\\
\label{U2}
\Upsilon_b\!\left(x+\frac{1}{b}\right)&=&\frac{\Gamma(\frac{x}{b})}{\Gamma(1-\frac{x}{b})}b^{-1+\frac{2x}{b}}\Upsilon_b(x).
\end{eqnarray}
Define $g(x)=b^{x^2-bx}\Upsilon_b(x)$.
Then, eq.~(\ref{U1}) is equivalent to the following:
$$
g(x+b)=\frac{b\Gamma(bx)}{\Gamma(1-bx)}g(x).
$$
Let us define $f(\sigma)=b^\sigma\Gamma(\sigma)g(b\sigma)$, where $\sigma=x/b$ is a fixed parameter.
Then, one gets
\begin{equation}\label{feq}
f(\sigma+1)=\frac{\Gamma(1+b^2\sigma)}{\Gamma(1-b^2\sigma)}f(\sigma).
\end{equation}
It can be shown that
\begin{equation}\label{RG}
\frac{\Gamma(1+b^2\sigma)}{\Gamma(1-b^2\sigma)}=
\exp\left[-2\gamma b^2\sigma-2\sum\limits_{n=1}^{\infty}\frac{\zeta(2n+1)}{2n+1}(b^2\sigma)^{2n+1}\right],
\end{equation}
where \(\gamma\) denotes the Euler--Mascheroni constant and \(\zeta(s)\) is the Riemann zeta function.
Indeed, to derive eq.~(\ref{RG}), one can rely on the following fact.
\begin{Prop}[Taylor series of \(\operatorname{Log}(\Gamma(z))\)]
The natural logarithm of the gamma function admits the following power series expansion:  
\begin{equation}\label{LG}
\operatorname{Log}(\Gamma(z)) = -\gamma (z-1) + \sum_{k=2}^\infty \frac{(-1)^k \zeta(k)}{k} (z-1)^k,
\end{equation} 
which is valid for all \( z \in \mathbb{C} \) such that \( |z-1| < 1 \).\footnote{See e.g.:
[\url{https://proofwiki.org/wiki/Taylor_Series_of_Logarithm_of_Gamma_Function}].}
\end{Prop}
\noindent
By substituting $z\to z+1$ and $z\to -z+1$ into (\ref{LG}), we obtain
\begin{eqnarray*}
	\operatorname{Log}(\Gamma(z+1)) &=& -\gamma z + \sum_{k=2}^\infty \frac{\zeta(k)}{k} (-z)^k,\\
	\operatorname{Log}(\Gamma(-z+1)) &=& \gamma z + \sum_{k=2}^\infty \frac{\zeta(k)}{k} z^k.
\end{eqnarray*}
Hence,
\begin{eqnarray}\label{LG2}
	\operatorname{Log}\frac{\Gamma(z+1)}{\Gamma(-z+1)}&=&
	-2\gamma z 
	+\sum_{k=2}^\infty \frac{\zeta(k)}{k} (-z)^k
	-\sum_{k=2}^\infty \frac{\zeta(k)}{k} z^k\nonumber
	\\
	&=&
	-2\gamma z -2\sum_{n=1}^\infty \frac{\zeta(2n+1)}{2n+1} z^{2n+1},
\end{eqnarray}
where we utilized the fact that the difference of the series is nonzero when $k$ is odd. 
Therefore, by setting $z=b^2\sigma$, one obtains eq.~(\ref{RG}).

Thus, by combining eqs.~(\ref{feq}) and (\ref{RG}), one gets 
\begin{equation}\label{feq2}
f(\sigma+1)=f(\sigma)
\exp\left[-2\gamma b^2\sigma-2\sum\limits_{n=1}^{\infty}\frac{\zeta(2n+1)}{2n+1}(b^2\sigma)^{2n+1}\right].
\end{equation}
A solution to eq.~(\ref{feq2}) can be constructed by adopting the ansatz:
\begin{equation}\label{ansatz}
f(\sigma)=f_{0}\exp\left[b^2\gamma\phi_{1}(\sigma)+\sum\limits_{n=1}^{\infty}b^{2(2n+1)}
\frac{\zeta(2n+1)}{2n+1}\phi_{2n+1}(\sigma)\right].
\end{equation}
Eq.~(\ref{feq2}) is satisfied, provided that
\begin{eqnarray*}
	\phi_{1}(\sigma+1)&=&\phi_{1}(\sigma)-2\sigma,\\
	\phi_{3}(\sigma+1)&=&\phi_{3}(\sigma)-2\sigma^3,\\
	\cdots &&\\
	\phi_{2n+1}(\sigma+1)&=&\phi_{2n+1}(\sigma)-2\sigma^{2n+1}.
\end{eqnarray*}
It is useful to introduce the generating function $F(\sigma)$ for the coefficients 
$\phi_{n}(\sigma)$, namely: 
\begin{equation}\label{Fs}
	F(\sigma)=\sum\limits_{n=0}^{\infty}\phi_{n}(\sigma)\frac{\eta^n}{n!}.
\end{equation}
Obviously, $F(\sigma)$ depends on $\eta$. From (\ref{Fs}) it turns out that
\begin{eqnarray*}
	\sum\limits_{n=0}^{\infty}\phi_{n}(\sigma+1)\frac{\eta^n}{n!}&=&
	\sum\limits_{n=0}^{\infty}\phi_{n}(\sigma)\frac{\eta^n}{n!}
	-2\sum\limits_{n=0}^{\infty}\frac{(\sigma\eta)^n}{n!}
	\\
	&=&
	\sum\limits_{n=0}^{\infty}\phi_{n}(\sigma)\frac{\eta^n}{n!}-2{\rm e}^{\eta\sigma}
\end{eqnarray*}
or 
\begin{equation}\label{Fs+1}
	F(\sigma+1)=F(\sigma)-2{\rm e}^{\eta\sigma}.
\end{equation}
The difference equation (\ref{Fs+1}) does not determine the generating
function uniquely. Indeed, if $F(\sigma)$ is a solution, then
\[
\widetilde{F}(\sigma)=F(\sigma)+P(\sigma)
\]
is also a solution, provided that
\begin{equation}\label{periodic}
	P(\sigma+1)=P(\sigma).
\end{equation}
To see this, substituting $\widetilde{F}(\sigma)$ into
eq.~(\ref{Fs+1}) gives
\[
F(\sigma+1)+P(\sigma+1)
=
F(\sigma)+P(\sigma)-2{\rm e}^{\eta\sigma}.
\]
Since $F(\sigma)$ already satisfies eq.~(\ref{Fs+1}), the remaining
condition is precisely (\ref{periodic}). Therefore, the solution
contains an arbitrary one-periodic ambiguity.

A particular solution of eq.~(\ref{Fs+1}) can be obtained using the
ansatz
\[
F_{\rm part}(\sigma)=C(\eta){\rm e}^{\eta\sigma}.
\]
Substitution into eq.~(\ref{Fs+1}) gives
\[
C(\eta){\rm e}^\eta=C(\eta)-2,
\]
and hence
\[
C(\eta)=-\frac{2}{{\rm e}^\eta-1}.
\]
Therefore,
\begin{equation}\label{Fpart}
	F_{\rm part}(\sigma)=-\frac{2{\rm e}^{\eta\sigma}}{{\rm e}^\eta-1}.
\end{equation}

The complete solution can be written as
\begin{equation}\label{Fgeneral}
	F(\sigma)=F_{\rm part}(\sigma)+P(\sigma),
	\qquad P(\sigma+1)=P(\sigma).
\end{equation}
The remaining ambiguity is fixed by requiring that the coefficients
$\phi_n(\sigma)$ in the expansion (\ref{Fs}) are polynomials in
$\sigma$, as in Thorn's construction~\cite{Thorn}. A periodic contribution expanded
in powers of $\eta$ would generate periodic polynomial terms. Since the
only periodic polynomials are constants, the ambiguity reduces to an
$\eta$-dependent constant,
\[
P(\sigma)=C_0(\eta).
\]
The normalization condition
\[
F(0)=0
\]
(which corresponds to $\phi_n(0)=0$ for all coefficients) gives
\[
C_0(\eta)=\frac{2}{{\rm e}^\eta-1}.
\]
Consequently,
\[
F(\sigma)=
-\frac{2{\rm e}^{\eta\sigma}}{{\rm e}^\eta-1}
+\frac{2}{{\rm e}^\eta-1},
\]
and finally
\begin{equation}\label{FF}
	F(\sigma)=
	-2\frac{{\rm e}^{\eta\sigma}-1}{{\rm e}^\eta-1}.
\end{equation}

Expanding this generating function in powers of $\eta$ yields the
polynomial coefficients $\phi_n(\sigma)$ entering the asymptotic
expansion of the $\Upsilon_b$-function. In particular,
\begin{eqnarray*}
	\phi_{1}(\sigma)&=&\sigma(1-\sigma),\\
	\phi_{3}(\sigma)&=&-\frac{1}{2}\sigma^2(1-\sigma)^2,\\
	\phi_{5}(\sigma)&=&-\frac{1}{6}\sigma^2(1-\sigma)^2(2\sigma^2-2\sigma-1).
\end{eqnarray*}

In summary, we find
\[
f(\sigma)=\Gamma(\sigma)\, b^{\,b^2\sigma^2-b^2\sigma+\sigma}\,\Upsilon_b(b\sigma).
\]
Adopting the ansatz~(\ref{ansatz}) and fixing the overall normalization to \(f_0=b\,\Upsilon_0\),
with \(\Upsilon_0:=\Upsilon_b'(0)\), we finally obtain the formula 
\begin{equation}\label{Ub}
\Upsilon_b(b\sigma)\,\sim\,b\Upsilon_{0}\frac{b^{b^2\sigma(1-\sigma)-\sigma}}{\Gamma(\sigma)}
\exp\left[b^2\gamma\phi_{1}(\sigma)+\sum\limits_{n=1}^{\infty}b^{2(2n+1)}
\frac{\zeta(2n+1)}{2n+1}\phi_{2n+1}(\sigma)\right],\qquad b\to 0,
\end{equation}
where the symbol $\sim$ indicates that the right-hand side is an
asymptotic expansion of $\Upsilon_b(b\sigma)$ for $b\to0$ with fixed
$\sigma$. The derivation is carried out for fixed values of $\sigma$
away from the poles of the intermediate Gamma-function factors. In
particular, for sufficiently small $b$ one has $|b^2\sigma|<1$, so that
the expansion of $\log\Gamma(1\pm b^2\sigma)$ used above is valid.
The resulting expression is then understood by analytic continuation
in $\sigma$, consistently with the meromorphic structure of the
underlying Liouville CFT quantities.

This completes the derivation of the small-$b$ asymptotic expansion of
$\Upsilon_b(b\sigma)$ that will be substituted into the DOZZ formula in
the next subsection to extract the coefficients $\Omega_n(\sigma_i)$.\footnote{
The functional relations used above are exact. The asymptotic character
of eq.~(\ref{Ub}) arises when the exact functional relation is solved
perturbatively in powers of $b^2$, following Thorn's analysis.}

\subsection{Semiclassical expansion of the DOZZ formula}
\label{smallb-DOZZ}
Substituting \(\alpha_i=b\sigma_i\) into the DOZZ expression (\ref{DOZZ}) and using the standard identity
$$
\Upsilon_{b}\!\left(x-\frac{1}{b}\right)=\gamma\left(\frac{x}{b}-\frac{1}{b^2}\right)^{-1}
b^{1+\frac{2}{b^2}-\frac{2x}{b}}\Upsilon_{b}(x)
$$
to trade the $\Upsilon_{b}(\sum_{i}b\sigma_i - Q)$ factor, the DOZZ constant can be rearranged into the form
\begin{eqnarray}\label{CC}
	C(b\sigma_1, b\sigma_2, b\sigma_3)&=& 
	\frac{\pi\tilde\mu\gamma(b^{-2})\left(\pi\mu\gamma(b^2)b^{-2b^2}\right)^{1-\sum_{i}\sigma_i}
		\gamma\left(\sum_{i}\sigma_i-1-\frac{1}{b^2}\right)}{b^{3}}\nonumber
	\\[5pt]
	&&\hspace{-100pt}\times\;
	\frac{\Upsilon_{0}\Upsilon_{b}(2b\sigma_1)\Upsilon_{b}(2b\sigma_2)\Upsilon_{b}(2b\sigma_3)}
	{\Upsilon_{b}(b(\sum_{i} \sigma_i - 1))
		\Upsilon_{b}(b(\sigma_1 + \sigma_2 -\sigma_3)) 
		\Upsilon_{b}(b(\sigma_2 + \sigma_3 - \sigma_1))
		\Upsilon_{b}(b(\sigma_3+\sigma_1-\sigma_2))},
\end{eqnarray}
where we use $\pi\tilde\mu\gamma(b^{-2})=\left(\pi\mu\gamma(b^{2})\right)^{b^{-2}}$ as before.

In the leading-order approximation as \(b\to 0\), the asymptotic expansion in~(\ref{Ub}) yields
\begin{equation}
\Upsilon_b(b\sigma)\sim
\Upsilon_0 \frac{b^{1-\sigma}}{\Gamma(\sigma)},
\qquad b\to 0.
\end{equation}
Substituting this asymptotic form into the ratio of \(\Upsilon\)-functions appearing in the DOZZ structure constant~(\ref{CC}), 
we obtain
\begin{eqnarray}\label{CCC}
	C(b\sigma_1, b\sigma_2, b\sigma_3)&\sim& 
	\frac{\pi\tilde\mu\gamma(b^{-2})\left(\pi\mu\gamma(b^2)b^{-2b^2}\right)^{1-\sum_{i}\sigma_i}
		\gamma\left(\sum_{i}\sigma_i-1-\frac{1}{b^2}\right)}{b^{5}}\nonumber
	\\[5pt]
	&&\hspace{-80pt}\times\;
	\frac{\Gamma(\sigma_{1}+\sigma_{2}+\sigma_{3}-1)
		\Gamma(\sigma_{2}+\sigma_{3}-\sigma_{1})\Gamma(\sigma_{3}+\sigma_{1}-\sigma_{2})\Gamma(\sigma_{1}+\sigma_{2}-\sigma_{3})}
	{\Gamma(2\sigma_{1})\Gamma(2\sigma_{2})\Gamma(2\sigma_{3})}
	\\[5pt]
	&=:& P(b;\sigma_1,\sigma_2,\sigma_3)\times A(\sigma_1,\sigma_2,\sigma_3).
\end{eqnarray}
This asymptotic behavior was first derived by Zamolodchikovs~\cite{ZZ5} using the path integral formulation and fixed-area 
techniques. It was later revisited in~\cite{HMW} by Harlow, Maltz, and Witten.

We now address the perturbative corrections to the zeroth-order result (\ref{CCC}).
Note that formula (\ref{Ub}) can be expressed as follows:
\begin{equation}\label{Ub2}
\Upsilon_b(b\sigma)\,\sim\,\left(\Upsilon_{0}\frac{b^{1-\sigma}}{\Gamma(\sigma)}\right)
b^{b^2(\sigma-\sigma^2)}{\rm e}^{\Phi(\sigma)},
\end{equation}
where
\begin{equation}\label{Phi}
\Phi(\sigma):=b^2\gamma\phi_{1}(\sigma)+\sum\limits_{n=1}^{\infty}b^{2(2n+1)}
\frac{\zeta(2n+1)}{2n+1}\phi_{2n+1}(\sigma).
\end{equation}
The factor on the right-hand side of (\ref{Ub2}), enclosed in parentheses, represents the ``classical contribution'' of each 
$\Upsilon_b$-function to (\ref{CCC}). Applying (\ref{Ub2}) to (\ref{CC}), we obtain
\begin{eqnarray}\label{CCC2}
	C(b\sigma_1, b\sigma_2, b\sigma_3)&\sim& 
	\frac{\pi\tilde\mu\gamma(b^{-2})\left(\pi\mu\gamma(b^2)b^{-2b^2}\right)^{1-\sum_{i}\sigma_i}
		\gamma\left(\sum_{i}\sigma_i-1-\frac{1}{b^2}\right)}{b^{5}}\nonumber
	\\[5pt]
	&&\hspace{-50pt}\times\;
	\frac{\Gamma(\sigma_{1}+\sigma_{2}+\sigma_{3}-1)
		\Gamma(\sigma_{2}+\sigma_{3}-\sigma_{1})\Gamma(\sigma_{3}+\sigma_{1}-\sigma_{2})\Gamma(\sigma_{1}+\sigma_{2}-\sigma_{3})}
	{\Gamma(2\sigma_{1})\Gamma(2\sigma_{2})\Gamma(2\sigma_{3})}\nonumber
	\\[5pt]
	&&\hspace{-50pt}\times\;
	\Omega(b;\sigma_1,\sigma_2,\sigma_3)
	\\[5pt]
	&=& P(b;\sigma_1,\sigma_2,\sigma_3)\times A(\sigma_1,\sigma_2,\sigma_3)\times 
	\Omega(b;\sigma_1,\sigma_2,\sigma_3).
\end{eqnarray}
where
\begin{eqnarray}\label{Corr}
	\Omega(b;\sigma_1,\sigma_2,\sigma_3)&:=&
	b^{-2b^2(\sigma_1+\sigma_2+\sigma_3-1)}\exp\left[
	\Phi(2\sigma_1)+\Phi(2\sigma_2)+\Phi(2\sigma_3)\nonumber
	\right.\\
	&&\left.
	-\Phi(\sigma_1+\sigma_2-\sigma_3)
	-\Phi(\sigma_1+\sigma_3-\sigma_2)\nonumber
	\right.\\
	&&\left.-\Phi(\sigma_2+\sigma_3-\sigma_1)
	-\Phi(\sigma_1+\sigma_2+\sigma_3-1)
	\right].
\end{eqnarray}
The factor $\Omega(b;\sigma_1,\sigma_2,\sigma_3)$ accounts for corrections to (\ref{CCC}). 
By incorporating successive terms in the expansion of $\Phi(\sigma)$, one can compute successive approximations 
to the expression (\ref{Corr}), thereby refining the corrections to (\ref{CCC}).
For instance, by considering only terms of order 
$b^2$ in the exponent of (\ref{Ub2}), i.e., 
$$
\Phi(\sigma)\sim b^2\gamma\phi_{1}(\sigma)=b^2\gamma\sigma(1-\sigma),
$$
one obtains from (\ref{Corr}):
\begin{eqnarray*}
	\Omega(b;\sigma_1,\sigma_2,\sigma_3)&\sim& b^{-2b^2(\sigma_1+\sigma_2+\sigma_3-1)}
	\exp\left[-2b^2\gamma(\sigma_1+\sigma_2+\sigma_3-1)\right]
	\\
	&=&{\rm e}^{-2b^2\gamma(\sigma_1+\sigma_2+\sigma_3-1)-2b^2(\sigma_1+\sigma_2+\sigma_3-1)\log b}.
\end{eqnarray*}
Taking the expansion up to order $b^6$ in the exponent, i.e.,  
$$
\Phi(\sigma)\sim b^2\gamma\sigma(1-\sigma)-b^6\frac{\zeta(3)}{6}\sigma^2(1-\sigma)^2
$$
yields 
\begin{eqnarray*}
	\Omega(b;\sigma_1,\sigma_2,\sigma_3)&\sim&
	b^{-2b^2(\sigma_1+\sigma_2+\sigma_3-1)}
	\exp\left[-2b^2\gamma(\sigma_1+\sigma_2+\sigma_3-1)\right.
	\\
	&&-b^{6}\frac{2}{3}\zeta(3) 
	(\sigma_1+\sigma_2+\sigma_3-1)
	\left[
	\sigma_{1}^{2} 
	(-3\sigma_2-3\sigma_3+1)
	+3\sigma_{1}^{3}\right.
	\\
	&&\left.
	+\sigma_1\left(-3\sigma_{2}^{2}
	+\sigma_2(6\sigma_3+2)
	+(2-3\sigma_3)\sigma_3-2\right)\right.
	\\
	&&\left.\left.
	+(\sigma_2+\sigma_3)
	\left(3\sigma_{2}^{2}-6\sigma_2\sigma_3
	+\sigma_2+3\sigma_{3}^{2}
	+\sigma_3\right)
	-2\sigma_2-2\sigma_3+1
	\right]\right].
\end{eqnarray*}
Successive approximations of the factor (\ref{Corr}), obtained by including higher-order terms in the expansion (\ref{Phi}), 
result in increasingly complex expressions. However, their calculation is straightforward with the use of Mathematica.

Expanding the exponential in the factor
$\Omega(b;\sigma_1,\sigma_2,\sigma_3)$ produces the corresponding asymptotic expansion 
of the DOZZ formula in the small-$b$ regime.
Truncating this expansion at a finite order gives
\begin{eqnarray}
C(b\sigma_1,b\sigma_2,b\sigma_3)
&\sim&
P(b;\sigma_1,\sigma_2,\sigma_3)\,
A(\sigma_1,\sigma_2,\sigma_3)\,
b^{-2b^2(\sigma_1+\sigma_2+\sigma_3-1)}
\nonumber\\
&\times&
\left[
\sum_{n=0}^{k-1} b^{2n}\,
\Omega_n(\sigma_1,\sigma_2,\sigma_3)
+\mathcal{O}(b^{2k})
\right],
\qquad b\to0,\quad \sigma_i\ {\rm fixed}.
\end{eqnarray}
Here, the combination
$P(b;\sigma_1,\sigma_2,\sigma_3)
A(\sigma_1,\sigma_2,\sigma_3)
b^{-2b^2(\sigma_1+\sigma_2+\sigma_3-1)}
=\mathcal{P}(b;\sigma_1,\sigma_2,\sigma_3)$
is exactly the prefactor~(\ref{Pre}) defined in the introduction,
encapsulating all explicit $b$- and $\sigma_i$-dependence, while 
the bracketed expression contains the asymptotic corrections organized
order by order in powers of $b^2$.\footnote{
	The resulting small-$b$ expansion of the DOZZ structure constant is
	initially obtained for fixed values of $\sigma_i$ away from the poles
	introduced by the intermediate Gamma-function factors. The asymptotic
	limit is taken as $b\to0$ with the $\sigma_i$ held fixed. The resulting
	expression is then understood for generic $\sigma_i$ by meromorphic
	analytic continuation, away from the poles of the DOZZ structure
	constant.}
For example, the first coefficients are
\begin{eqnarray}
\Omega_0(\sigma_1,\sigma_2,\sigma_3) &=& 1,
\\[5pt]
\Omega_1(\sigma_1,\sigma_2,\sigma_3) &=&-2\gamma(\sigma_1+\sigma_2+\sigma_3-1),
\\[5pt]
\Omega_2(\sigma_1,\sigma_2,\sigma_3) &=&2\gamma^2(\sigma_1+\sigma_2+\sigma_3-1)^2.
\end{eqnarray}
and
\begin{eqnarray}
\Omega_3(\sigma_1,\sigma_2,\sigma_3) &=& \frac{1}{6} \Bigg( 
-8\gamma^3(\sigma_1+\sigma_2+\sigma_3-1)^3 
\nonumber \\[5pt]
&& \quad -4(\sigma_1+\sigma_2+\sigma_3-1)\Big( 
1 + 3\sigma_1^3 - 2\sigma_2 - 2\sigma_3 
\nonumber \\[5pt]
&& \quad + \sigma_1^2(1 - 3\sigma_2 - 3\sigma_3) 
+ \sigma_1\big(-2+(2 - 3\sigma_2)\sigma_2 
\nonumber \\[5pt]
&& \qquad + (2 + 6\sigma_2)\sigma_3 - 3\sigma_3^2 \big) 
\nonumber \\[5pt]
&& \quad + (\sigma_2 + \sigma_3)\left( \sigma_2 + 3\sigma_2^2 + \sigma_3 
- 6\sigma_2\sigma_3 + 3\sigma_3^2 \right) 
\Big)\zeta(3) \Bigg).
\end{eqnarray}

The DOZZ three-point function is symmetric under permutations of the insertions \(\alpha_1,\alpha_2,\alpha_3\). 
Accordingly, the coefficients \(\Omega_n(\sigma_i)\) appearing in the small-\(b\) expansion should respect the same symmetry. 
(The prefactor \(\mathcal{P}(b;\sigma_i)\) is manifestly symmetric; see (\ref{Pre}).)
For \(\Omega_1(\sigma_i)\) and \(\Omega_2(\sigma_i)\) the permutation symmetry in the parameters 
\(\sigma_i\) is manifest, while for \(\Omega_3(\sigma_i)\) the symmetry can be verified by direct computation. 
Moreover, looking at the structure of \(\Omega(b;\sigma_1,\sigma_2,\sigma_3)\) in eq.~(\ref{Corr}), it is straightforward 
to see that each coefficient \(\Omega_n(\sigma_i)\) is a symmetric polynomial of the variables \(\sigma_i\). 
Indeed, the argument of the exponential in \eqref{Corr} is expressed in terms of the function \(\Phi\). 
The first three terms depend on \(2\sigma_1,2\sigma_2,2\sigma_3\) and are therefore symmetric under permutations of the 
\(\sigma_i\). The remaining four terms are exchanged among themselves under any transposition of two \(\sigma_i\).
Successive approximations of the function \(\Phi\) are polynomials in its argument. 
Since the exponential in \eqref{Corr} is constructed from these approximants, 
it inherits their polynomial dependence; consequently, the polynomial structure is preserved.

\section{Perturbative Liouville expansion and celestial holography}
\label{perturbative-liouville}
In this section we discuss a possible perturbative interpretation of the
small-$b$ expansion of the DOZZ three-point function for light operators
in the context of celestial holography
\cite{Stieberger2023a,Stieberger2023b,Melton:2024akx}.
(See also related discussions in~\cite{Giribet:2024vnk,Donnay:2025yoy}.)\footnote{For 
an introduction to celestial holography, see
\cite{Pasterski:2021rjz,Raclariu:2021zjz,Donnay:2023celestial,
Pasterski:2021dqe,Pasterski:2017kqt,Schreiber:2017jsr,Kalyanapuram:2020aya,Banerjee:2020kaa}.}
Rather than presenting a complete derivation of celestial loop amplitudes,
our aim is to identify the structures in Liouville theory that may provide
a systematic framework for organizing quantum corrections. We discuss
three complementary viewpoints: 
the minisuperspace (zero-mode) reduction
interpreted as Liouville quantum mechanics (LQM),
the path-integral expansion around
classical configurations, and the operator
perturbative bootstrap. Together, these approaches suggest a possible
route towards constraining finite-$b$ corrections to celestial
correlators.

The basic observation is that Liouville conformal field theory provides a
two-dimensional framework for organizing celestial correlators associated
with four-dimensional gluon scattering in specific dilaton backgrounds
\cite{Stieberger2023a,Stieberger2023b,Melton:2024akx}.
In the light-operator limit
\[
\alpha_i=b\,\sigma_i,\qquad b\to0,\qquad \sigma_i\ {\rm fixed},
\]
the DOZZ three-point function admits the small-$b$ asymptotic expansion
(\ref{dozzlight})--(\ref{Pre}),
where
\begin{equation}
	A(\sigma_1,\sigma_2,\sigma_3)=
	\frac{\Gamma(\sigma_{1}+\sigma_{2}+\sigma_{3}-1)
		\Gamma(\sigma_{2}+\sigma_{3}-\sigma_{1})
		\Gamma(\sigma_{3}+\sigma_{1}-\sigma_{2})
		\Gamma(\sigma_{1}+\sigma_{2}-\sigma_{3})}
	{\Gamma(2\sigma_{1})\Gamma(2\sigma_{2})\Gamma(2\sigma_{3})}
	\label{eq:classical-prefactor-eng}
\end{equation}
is the classical Gamma-function prefactor, while the functions
$\Omega_n(\sigma_i)$ describe the successive quantum corrections in the
small-$b$ expansion.

The classical prefactor represents the leading contribution surviving in
the semiclassical limit. From the celestial perspective, it provides the
tree-level building block for the perturbative expansion, encoding the
kinematic dependence of the corresponding three-point celestial
coupling. The interpretation of the higher-order coefficients
$\Omega_n(\sigma_i)$ as loop corrections is understood here as a proposed
identification, whose precise relation to celestial amplitudes requires
additional dynamical input, including the Mellin transform, current
sector, renormalization prescription, and comparison with explicit loop
amplitude calculations.

A particularly transparent interpretation of the leading factor
$A(\sigma_1,\sigma_2,\sigma_3)$ is provided by the minisuperspace
reduction of Liouville theory. Restricting the Liouville field to its
spatially constant mode,
\[
\varphi(x)=\varphi_0,
\]
reduces the theory to an effective one-dimensional quantum mechanics
with Hamiltonian
\begin{equation}
	H_{\mathrm{LQM}}
	=
	-\frac{1}{2}\frac{{\rm d}^{2}}{{\rm d}\varphi_0^{2}}
	+\mu\,{\rm e}^{2b\varphi_0},
\end{equation}
whose eigenfunctions are expressed in terms of modified Bessel functions,
and are naturally parametrized by the Liouville momentum variables.

In this formulation, Liouville exponential insertions are represented by
operators acting on the quantum-mechanical Hilbert space. The relevant
quantities are therefore matrix elements of exponential vertex operators
between Liouville quantum-mechanical states, of the schematic form
\begin{equation}
	\left\langle \Psi_{\sigma_1}\right|
	{\rm e}^{2b\sigma_2\varphi_0}
	\left|\Psi_{\sigma_3}\right\rangle .
\end{equation}
Such matrix elements arise naturally in the canonical weak-coupling
treatment of Liouville theory, where the zero mode is treated exactly as
Liouville quantum mechanics and the nonzero modes are treated
perturbatively~\cite{Thorn}. This provides the standard minisuperspace
interpretation of the leading light-operator contribution. The resulting
matrix elements exhibit the same Gamma-function structures that appear in
$A(\sigma_1,\sigma_2,\sigma_3)$, motivating the interpretation of $A$ as
the leading zero-mode contribution to the three-point data.

The higher-order terms in the small-$b$ expansion admit a complementary
path-integral interpretation. Expanding the Liouville field around a
background configuration,
\[
\varphi=\varphi_{\rm cl}+\delta\varphi,
\]
gives
\begin{equation}
	S[\varphi]
	=
	S_{\rm cl}
	+\frac{1}{2}
	\langle\delta\varphi,\mathcal{D}\,\delta\varphi\rangle
	+S_{\rm int}[\delta\varphi],
\end{equation}
where $\mathcal{D}$ is the fluctuation operator and
$S_{\rm int}$ contains the interaction vertices for the fluctuation
field.

The connected quantum contributions are naturally organized in the
logarithm of the multiplicative correction factor. We therefore write
schematically
\begin{equation}
	\log \Omega(b;\sigma_i)
	=
	\sum_{n\geq1} b^{2n}\,{\Gamma}_n(\sigma_i),
\end{equation}
where $\Gamma_n$ denotes the connected contribution at order $b^{2n}$.
In this organization, the one-loop fluctuation determinant contributes to
$\Gamma_1$, while higher $\Gamma_n$ receive contributions from connected
higher-order diagrams, together with the corresponding counterterms and
normalization factors. The coefficients $\Omega_n$ in the expansion of
the full multiplicative factor $\Omega(b;\sigma_i)$ are obtained after
exponentiating the connected expansion and therefore are not, in general,
identical to individual loop contributions.

For example, the first two coefficients satisfy
\begin{equation}
\Omega_1=\Gamma_1,
\qquad
\Omega_2=\Gamma_2+\frac{1}{2}\Gamma_1^2.
\end{equation}
Thus, the relation
$\Omega_2=\frac{1}{2}\Omega_1^2$ found in the present calculation implies
\(\Gamma_2=0\). In particular, there is no independent connected contribution at order
$b^4$ in $\log\Omega$. This distinction is essential when relating the
small-$b$ coefficients to a diagrammatic loop expansion.

Within the operator formalism, the coefficients $\Omega_n(\sigma_i)$
represent finite-order corrections to the Liouville CFT data. Expanding
three- and four-point functions in powers of $b^2$ and imposing crossing
symmetry order by order leads to a perturbative bootstrap problem. At a
given order, the crossing equations constrain combinations of corrections
to OPE coefficients, conformal dimensions, and the spectral measure.

In this sense, the perturbative bootstrap provides consistency conditions
for the finite-order coefficients obtained from the small-$b$ expansion.
It can also help distinguish contributions fixed by CFT consistency from
those that depend on normalization or renormalization conventions and
therefore require additional physical input.

The relation between the finite-$b$ Liouville expansion and celestial
Yang--Mills amplitudes is more properly viewed as a perturbative research
program. The present calculation determines the small-$b$ Liouville
three-point data, but does not by itself establish a direct correspondence
between the coefficients $\Omega_n$ and loop amplitudes.

A complete construction would require, in particular, the implementation
of the appropriate Mellin transform, the treatment of the celestial
current sector, a consistent renormalization prescription, and the
verification of unitarity and factorization properties. Where explicit
bulk amplitude results are available, these ingredients should also allow
for a direct comparison with known tree-level and loop-level amplitudes.
Such comparisons would provide a test of the proposed Liouville
description and determine any remaining normalization or renormalization
ambiguities.

Taken together, the minisuperspace, path-integral, and
operator/bootstrap viewpoints provide complementary ways of organizing
the finite-$b$ corrections to Liouville three-point data. The leading
factor $A(\sigma_i)$ has a natural zero-mode interpretation, while the
higher-order coefficients $\Omega_n(\sigma_i)$ encode successive terms in
the small-$b$ expansion. Their interpretation in terms of celestial loop
corrections requires the additional steps described above. We therefore
regard the present results as Liouville input for a perturbative celestial
program rather than as a complete construction of celestial
Yang--Mills loop amplitudes.

\section{Conclusions}
\label{conclusions}
In this work we developed a systematic asymptotic small-$b$ expansion of
the DOZZ three-point structure constant for light Liouville operators.
Using the small-$b$ asymptotics of the $\Upsilon_b$-function, we obtained
closed-form expressions for the first few coefficients and verified their
permutation symmetry. The expansion is understood asymptotically for
$b\to0$ with fixed light parameters, with analytic continuation used to
extend the resulting expressions to generic values of the parameters.

We also discussed several possible interpretations of the resulting
finite-$b$ Liouville data. The leading contribution admits a
minisuperspace interpretation in terms of matrix elements of exponential
Liouville operators between quantum-mechanical states. At higher orders,
the distinction between the multiplicative coefficients $\Omega_n$ and
the connected coefficients appearing in $\log\Omega$ becomes essential.
The latter provide the natural quantities for a diagrammatic
interpretation of connected quantum corrections.

The connection to celestial holography is formulated here as a perturbative
program. The present calculation provides finite-$b$ Liouville data that
may serve as input for such a construction, but a complete derivation of
celestial Yang--Mills loop amplitudes requires additional ingredients,
including the Mellin transform, the celestial current sector,
renormalization, unitarity and factorization checks, and comparison with
explicit amplitude calculations.

A concrete implementation of the perturbative Liouville program is
presented in ref.~\cite{FPP2026}. There, a refined Mellin--Liouville
prescription is developed for the celestial three-gluon amplitude,
resolving an ambiguity in the original proposal and relating the
Liouville parameters to the celestial Mellin variables. The present work
provides the small-$b$ Liouville data that enter this construction.

\appendix
\section{Three- and four-point DOZZ correlators on the Riemann sphere}
\renewcommand{\theequation}{A.\arabic{equation}}
\setcounter{equation}{0}
\label{app1}
Quantum Liouville theory on the Riemann sphere
$\widehat{\mathbb{C}}=\mathbb{C}\cup\{\infty\}$
has a physical Hilbert space given by a continuous direct integral of
Virasoro representations,
\begin{equation}
	{\cal H}_{L}
	=
	\int\limits_{0}^{\infty} {\rm d}P\;
	{\cal V}_{h(P)}\otimes{\cal V}_{h(P)},
	\qquad
	\alpha(P)=\frac{Q}{2}+iP,
\end{equation}
where
\[
h(P)=\alpha(P)\bigl(Q-\alpha(P)\bigr)
=\frac{Q^2}{4}+P^2.
\]
The identity representation with $h=0$ is not part of the normalizable
Liouville spectrum. Accordingly, we do not introduce an
$SL(2,\mathbb C)$-invariant vacuum as an element of ${\cal H}_{L}$.
Sphere correlation functions will instead be defined directly as
correlation functions of Liouville primary fields.

For generic Liouville momenta, conformal invariance fixes the coordinate
dependence of the three-point function up to the structure constant,
\begin{eqnarray}\label{3DOZZ}
	\left\langle
	{\sf V}_{\alpha_1}(z_1,\bar z_1)
	{\sf V}_{\alpha_2}(z_2,\bar z_2)
	{\sf V}_{\alpha_3}(z_3,\bar z_3)
	\right\rangle &=&\nonumber
	\\[5pt]
	&&\hspace{-190pt}=
	|z_{1}-z_{2}|^{2(h_{3}-h_{1}-h_{2})}
	|z_{1}-z_{3}|^{2(h_{2}-h_{1}-h_{3})}
	|z_{2}-z_{3}|^{2(h_{1}-h_{2}-h_{3})} 
	C(\alpha_{1},\alpha_{2},\alpha_{3}),
\end{eqnarray}
where $C(\alpha_1,\alpha_2,\alpha_3)$ is the DOZZ structure constant
given in eq.~(\ref{DOZZ}). For momenta outside the normalizable spectrum,
including the light momenta $\alpha_i=b\sigma_i$, this expression is
understood by meromorphic analytic continuation.

The DOZZ Liouville $s$-channel four-point function is expressed for the standard locations
$z_1 = \infty, z_2 = 1, z_3 = x, z_4 = 0$
as an integral over the continuous spectrum:
\begin{eqnarray}
	\label{four:point:} && \hspace*{-1.5cm} \Big\langle
	\textsf{V}_{\alpha_1}(\infty,\infty)\textsf{V}_{\alpha_2}(1,1)
	\textsf{V}_{\alpha_3}(x,\bar x)\textsf{V}_{\alpha_4}(0, 0)
	\Big\rangle =
	\\
	\nonumber && \int\limits_{\frac{Q}{2} + i{\mathbb
			R}_{\geq 0}}\!\!\!\!\textrm{d}\alpha\; 
	C(\alpha_1,\alpha_2,\alpha)C(Q-\alpha,\alpha_3,\alpha_4)
	\left| 
	{\cal F}_{c,h_s}\!\left[_{h_{1}\;h_{4}}^{h_{2}\;h_{3}}\right]\!(x)\right|^2.
\end{eqnarray}
Here ${\cal F}_{c,h_s}\!\left[_{h_{1}\;h_{4}}^{h_{2}\;h_{3}}\right]\!(x)$ 
is the four-point Virasoro conformal block. To define this universal special function of 2d CFT 
let us consider the operator product expansion of primary fields:
\begin{equation}
	\phi_{h_1,\bar h_1}(x,\bar {x})\phi_{h_2,\bar h_2}(0,0)=\sum_{s}
	C_{12s}
	\,x^{h_{s}-h_{1}-h_{2}}
	\overline{x}^{\bar{h}_{s}-\bar{h}_{1}-\bar{h}_{2}}
	\Psi_{12s}(x,\bar x),
\end{equation}
where for each $s$ the descendant contributions
$\Psi_{12s}(x,\bar x)$ are fixed by the conformal Ward identities.
More precisely, the descendant states entering the conformal block
belong to the tensor product
${\cal V}_{c,h_s}\otimes{\cal V}_{c,\bar h_s}$
of the holomorphic and antiholomorphic Verma modules with highest
weights $h_s$ and $\bar h_s$, respectively.\footnote{Here, $h_s=\bar h_s$.}
The relative coefficients of the descendant states, and hence their
$x$-dependence, are uniquely fixed by the conformal Ward identities.
In the left sector one has
\begin{equation}\label{psi12s}
	|\,\psi_{12s}(x)\,\rangle \;=\;\left|\,h_s\,\right\rangle + \sum\limits_{n=1}^\infty x^n
	\,\left|\,\beta^{n}_{c,h_{s}}\!\left[\,^{h_2}_{h_1}\right]\,\right\rangle,
\end{equation}
where $\left|\,h_s\,\right\rangle$ is the highest weight vector in
${\cal V}_{c,h_{s}}$ and
$\left|\,\beta^{n}_{ c,h_{s}}\!\left[\,^{h_2}_{h_1}\right]\,\right\rangle
\in {\cal V}_{c,h_{s}}^n\subset {\cal V}_{c,h_{s}}$.
In the $n$-level subspace ${\cal V}^{n}_{h_s}$ we shall use  a standard basis
consisting of vectors of the form
\begin{equation}
	\label{basis}
	|\,h^{n}_{I}\,\rangle=L_{-I}|\,h\,\rangle 
	:= L_{-i_{k}}\ldots L_{-i_{2}}L_{-i_{1}}|\,h\,\rangle,
\end{equation}
where $I=\lbrace
i_{k},\ldots ,i_{1}\rbrace$ is an ordered ($i_{k}\geq \ldots\geq
i_{1}\geq 1$) sequence of positive integers of the length $|I|:=i_{k}+\ldots+i_{1}=n$.
The conformal Ward identity for the three-point function implies the equations
$$
L_i\left|\,\beta^{n}_{c,h_{s}}\!\left[\,^{h_2}_{h_1}\right]\,\right\rangle
= (h_{s} + ih_1 -h_2 + n-i)
\left|\,\beta^{n-i}_{c,h_{s}}\!\left[\,^{h_2}_{h_1}\right]\,\right\rangle
$$
which in the basis (\ref{basis}) take the form
\begin{equation}
	\label{betas}
	\sum\limits_{|J|=n} \Big[ G_{c,h_{s}}^{n}\Big]_{IJ}
	\beta^{n}_{c,h_{s}}\!\left[\,^{h_2}_{h_1}\right]^J
	\;=\;
	\gamma_{h_{s}}\!\left[\,^{h_2}_{h_1}\right]_I\,,
\end{equation}
where 
\begin{equation}\label{Gram}
	\Big[G_{c,h}^{n}\Big]_{IJ}\;=\;\langle\,h_{I}^{n}\,|\,h_{J}^{n}\,\rangle
\end{equation}
is the Gram matrix and
\begin{eqnarray}
	\label{gamma}
	\gamma_{h_{s}}\!\left[\,^{h_2}_{h_1}\right]_I\!\! &=&\!\!
	(h_{s} +i_k h_1 -h_2 + i_{k-1} \!+\!\dots \!+\!i_1)\times \dots\\
	\nonumber
	&\dots &\times(h_{s} +i_2 h_1 -h_2 +i_1) (h_{s} +i_1 h_1-h_2 ) \ .
\end{eqnarray}
For all  values of the variables $c$ and $h_{s}$ for which the Gram matrices are invertible
the equations (\ref{betas}) admit  unique solutions
$$
\beta^{n}_{c,h_{s}}\!\left[\,^{h_2}_{h_1}\right]^I
=
\sum\limits_{|J|=n} \Big[ G_{c,h_{s}}^{n}\Big]^{IJ}
\gamma_{h_{s}}\!\left[\,^{h_2}_{h_1}\right]_J\,.
$$
In this range the four-point conformal block defined as a product of vectors (\ref{psi12s}) 
has for fixed positions $\infty,1,x,0$ a formal power series representation
\cite{BPZ}:
\begin{eqnarray}
	\label{block}
	{\cal F}_{c,h_s}\!\left[_{h_{1}\;h_{4}}^{h_{2}\;h_{3}}\right]\!(x)
	&=&
	x^{h_{s}-h_{3}-h_{4}}\left( 1 +
	\sum_{n=1}^\infty x^{n}
	{\cal F}^{(n)}_{c,h_{s}}\!\left[_{h_{1}\;h_{4}}^{h_{2}\;h_{3}}\right]\right),
	\\
	\label{blockBn}
	{\cal F}^{(n)}_{c,h_{s}}\!\left[_{h_{1}\;h_{4}}^{h_{2}\;h_{3}}\right]
	&=&
	\sum\limits_{|I|=|J|=n}
	\gamma_{h_{s}}\!\left[\,^{h_2}_{h_1}\right]_I
	\Big[G_{c,h_{s}}^{n}\Big]^{IJ}
	\gamma_{h_{s}}\!\left[\,^{h_3}_{h_4}\right]_J.
\end{eqnarray}

\section*{Acknowledgments}
MRP acknowledges Veronika Piatek for her unwavering support during the preparation of this work.
This research was funded by National Science Centre, Poland (NCN) under grant no.
2023/49/B/ST2/01371.


\begin{thebibliography}{99}
\providecommand{\href}[2]{#2}
\providecommand{\arxivref}[2]{\href{http://arxiv.org/abs/#1}{#2}}
\providecommand{\doiref}[2]{\href{http://dx.doi.org/#1}{#2}}
\providecommand{\nbbstauthor}[1]{#1}
\providecommand{\nbbstjournal}[1]{\textsf{#1}}
\providecommand{\nbbsttitle}[1]{\textit{#1}}
\providecommand{\nbbsturl}[1]{\texttt{#1}}
\providecommand{\nbbsteprint}[1]{\texttt{#1}}
\providecommand{\nbbststyle}{\raggedright\small\parskip0pt}
\nbbststyle

\bibitem{BPZ}
\nbbstauthor{A.A.~Belavin, A.M.~Polyakov, A.B.~Zamolodchikov}, 
\nbbsttitle{Infinite conformal symmetry in 2D quantum field theories}, 
\nbbstjournal{Nucl.~Phys.~B241 (1984) 333}.

\bibitem{S}
\nbbstauthor{N.~Seiberg}, 
\nbbsttitle{Notes on Quantum Liouville Theory and Quantum Gravity}, 
\nbbstjournal{Prog.~Theor.~Phys.~Suppl.~102 (1990) 319-349}.

\bibitem{DO} 
\nbbstauthor{H.~Dorn, H.J.~Otto}, 
\nbbsttitle{Two and three point functions in Liouville theory}, 
\nbbstjournal{Nucl.~Phys.~B429 (1994) 375-388, hep-th/9403141}.

\bibitem{T1} 
\nbbstauthor{J.~Teschner}, 
\nbbsttitle{On the Liouville three point function}, 
\nbbstjournal{Phys.~Lett.~B363 (1995) 65-70, hep-th/9507109}.

\bibitem{ZZ5}
\nbbstauthor{A.B.~Zamolodchikov, Al.B.~Zamolodchikov},
\nbbsttitle{Structure constants and conformal bootstrap in Liouville field theory},
\nbbstjournal{Nucl.~Phys.~B477 (1996) 577-605, hep-th/9506136}.

\bibitem{T3}
\nbbstauthor{J.~Teschner},
\nbbsttitle{Liouville theory revisited}, 
\nbbstjournal{Class.~Quant.~Grav.~18 (2001) R153-R222, hep-th/0104158}.

\bibitem{T4}
\nbbstauthor{J.~Teschner},
\nbbsttitle{A lecture on the Liouville vertex operators}, 
\nbbstjournal{Int.~J.~Mod.~Phys.~A19 (2004) 436-458, hep-th/0303150}.

\bibitem{Nakayama:2004vk}
\nbbstauthor{Y.~Nakayama},
\nbbsttitle{Liouville field theory: A Decade after the revolution}
\nbbstjournal{Int.~J.~Mod.~Phys.~A19 (2004) 2771-2930, hep-th/0402009}.

\bibitem{Stieberger2023a}
\nbbstauthor{S.~Stieberger, T.R.~Taylor, B.~Zhu}, 
\nbbsttitle{Celestial Liouville theory for Yang--Mills amplitudes}, 
\nbbstjournal{Phys.~Lett.~B836 (2023) 137588}.

\bibitem{Stieberger2023b}
\nbbstauthor{S.~Stieberger, T.R.~Taylor, B.~Zhu}, 
\nbbsttitle{Yang--Mills as a Liouville theory}, 
\nbbstjournal{Phys.~Lett.~B846 (2023) 138229}.

\bibitem{Melton:2024akx}
\nbbstauthor{W.~Melton, A.~Sharma, A.~Strominger, T.~Wang},
\nbbsttitle{Celestial Dual for Maximal Helicity Violating Amplitudes},
\nbbstjournal{Phys.~Rev.~Lett.~133 (2024) 9, 091603, arXiv:2403.18896}.

\bibitem{HJ04} 
\nbbstauthor{L.~Hadasz, Z.~Jask\'{o}lski}, 
\nbbsttitle{Classical Liouville action on the sphere with three hyperbolic singularities}, 
\nbbstjournal{Nucl.~Phys.~B694 (2004) 493, hep-th/0309267}. 

\bibitem{HJP05}
\nbbstauthor{L.~Hadasz, Z.~Jask\'{o}lski, M.~Pi\c{a}tek}, 
\nbbsttitle{Classical geometry from the quantum Liouville theory},
\nbbstjournal{Nucl.~Phys.~B724 (2005) 529, hep-th/0504204}.

\bibitem{Thorn}
\nbbstauthor{Ch.B.~Thorn},
\nbbsttitle{Liouville Perturbation Theory},	
\nbbstjournal{Phys.~Rev.~D66 (2002) 027702, arXiv:hep-th/0204142}.

\bibitem{Perlmutter:2015} 
\nbbstauthor{E.~Perlmutter},
\nbbsttitle{Virasoro conformal blocks in closed form},
\nbbstjournal{JHEP 02 (2015) 023, arXiv:1502.07742}.

\bibitem{Bombini:2018jrg}
\nbbstauthor{A.~Bombini, S.~Giusto, R.~Russo},  
\nbbsttitle{A note on the Virasoro blocks at order $1/c$}, 
\nbbstjournal{Eur.~Phys.~J.~C79, 3 (2019), arXiv:1807.07886}.

\bibitem{HMW}
\nbbstauthor{D.~Harlow, J.~Maltz, E.~Witten},
\nbbsttitle{Analytic Continuation of Liouville Theory},
\nbbstjournal{JHEP 1112 (2011) 071, arXiv:1108.4417}.	

\bibitem{Giribet:2024vnk}  
\nbbstauthor{G.~Giribet}, 
\nbbsttitle{Remarks on celestial amplitudes and Liouville theory},
\nbbstjournal{Int.~J.~Mod.~Phys.~D34 (2025) 01, arXiv:2403.03374}.

\bibitem{Donnay:2025yoy}
\nbbstauthor{L.~Donnay, G.~Giribet, B.~Valsesia},
\nbbsttitle{MHV leaf amplitudes from parafermions},
\nbbstjournal{JHEP 06 (2025) 234, arXiv:2501.19332}.

\bibitem{Pasterski:2021rjz} 
\nbbstauthor{S.~Pasterski, M.~Pate, A.~M.~Raclariu},
\nbbsttitle{Celestial Holography},
\nbbstjournal{arXiv:2111.11392}.

\bibitem{Raclariu:2021zjz}
\nbbstauthor{A.~M.~Raclariu},
\nbbsttitle{Lectures on Celestial Holography},
\nbbstjournal{arXiv:2107.02075}.

\bibitem{Donnay:2023celestial}
\nbbstauthor{L.~Donnay},
\nbbsttitle{Celestial Holography: An Asymptotic Symmetry Perspective},
\nbbstjournal{Phys.~Rept.~1073 (2024) 1-41, arXiv:2310.12922}.

\bibitem{Pasterski:2021dqe}
\nbbstauthor{S.~Pasterski},
\nbbsttitle{Lectures on celestial amplitudes},
\nbbstjournal{Eur.~Phys.~J.~C81 (2021) 12, 1062, arXiv:2108.04801}.

\bibitem{Pasterski:2017kqt}
\nbbstauthor{S.~Pasterski, S.~H.~Shao, A.~Strominger},
\nbbsttitle{Flat Space Amplitudes and Conformal Symmetry of the Celestial Sphere},
\nbbstjournal{Phys.~Rev.~D96 (2017) 065026, arXiv:1701.00049}.

\bibitem{Schreiber:2017jsr}
\nbbstauthor{A.~Schreiber, A.~Volovich, M.~Zlotnikov},
\nbbsttitle{Tree-level gluon amplitudes on the celestial sphere},
\nbbstjournal{Phys.~Lett.~B781 (2018) 349-357, arXiv:1711.08435}.

\bibitem{Kalyanapuram:2020aya}
\nbbstauthor{N.~Kalyanapuram},
\nbbsttitle{Gauge and Gravity Amplitudes on the Celestial Sphere},
\nbbstjournal{Phys.~Rev.~D103 (2021) 085015, arXiv:2012.04579}.

\bibitem{Banerjee:2020kaa}
\nbbstauthor{S.~Banerjee, S.~Ghosh, P.~Paul},
\nbbsttitle{MHV Graviton Scattering Amplitudes and Current Algebra on the Celestial Sphere},
\nbbstjournal{JHEP 02 (2021) 176, arXiv:2008.04330}.

\bibitem{FPP2026}
\nbbstauthor{G.~Biskowski, F.~Ferrari, M.R.~Pi\c{a}tek, A.R.~Pietrykowski},
\nbbsttitle{A perturbative Liouville prescription for the celestial three-gluon amplitude},
\nbbstjournal{arXiv:2604.27736}.

\end{thebibliography}
\end{document}